\documentclass[twocolumn,showpacs,preprintnumbers,amsmath,amssymb]{revtex4}

\usepackage{graphicx}
\usepackage{dcolumn}
\usepackage{bm}
\usepackage{multirow}

\begin{document}

\title{Group theory analysis of electrons and phonons in N-layer graphene systems}

\author{L. M. Malard, D. L. Mafra, M. H. D. Guimar\~{a}es, M. S. C. Mazzoni and A. Jorio}
\address{Departamento de F\'{\i}sica, Universidade Federal de Minas
Gerais, 30123-970, Belo Horizonte, Brazil}

\date{\today}

\begin{abstract}

In this work we study the symmetry properties of electrons and
phonons in graphene systems as function of the number of layers. We
derive the selection rules for the electron-radiation and for the
electron-phonon interactions at all points in the Brillouin zone. By
considering these selection rules, we address the double resonance
Raman scattering process. The monolayer and bilayer graphene in the
presence of an applied electric field are also discussed.

\end{abstract}

\maketitle

\section{Introduction}

The current interest on graphene and its multilayered materials has
been stimulated by various experimental and theoretical works
addressing the physics of Dirac fermions and the potential for
device applications (\cite{geim07,castroreview} and references
therein). Group theory is a powerful theoretical tool to determine
eigenvectors, the number and the degeneracies of eigenvalues and to
obtain and understand the selection rules governing, for exemple,
electron-radiation and electron-phonon interactions. Although the
symmetry aspects of mono-layer graphene and graphite have been
largely discussed in the literature \cite{adobook}, the recent
findings generate interest in a group theory analysis depending on
the number of graphene layers.

This work presents a group theory analysis for electrons and phonons
in mono-, bi- and tri-layer graphene, extending for N layers
depending if N is even or odd. The selection rules for
electron-radiation interaction within the dipole approximation and
for electron scattering by phonons are derived. With these selection
rules, we discuss the double-resonance Raman (DRR) scattering
process, which has been widely used to characterize the number of
layers \cite{ferrari,gupta06,graf07} and to probe their electronic
and vibrational properties \cite{lmalard07,ni2008,bob07}. Finally,
we also discuss the differences when mono- and bi-layer graphene are
exposed to external electrical fields, giving insight on the gap
opening in the biased bilayer graphene
\cite{ohta06,neto07biased,mccann2006,johan07biased} and different
selection rules for the electron-phonon scattering (EPS) process.

Section \ref{sec2} gives the symmetry properties for monolayer and N
layer graphene depending if N is even or odd. The monolayer and
bilayer graphene are also considered in the presence of an electric
field perpendicular to the graphene plane. The notation adopted is
related to the space group symmetry, and conversion to the point
group notation can be found in the appendix\ref{appendix}. Section
\ref{sec3} presents the selection rules for the electron-radiation
interaction. Section \ref{sec4} shows the $\Gamma$ point Raman and
infrared active modes, and we extend the electron-phonon selection
rules to points in the interior of the Brillouin zone in section
\ref{sec5}. Considering both sections \ref{sec3} and \ref{sec5} we
address the DRR process for mono-, bi- and tri-layer graphene in
section \ref{sec6}. The main findings are summarized in section
\ref{sec7}.

\section{Symmetry properties}\label{sec2}

\subsection{Group of wavevector}\label{sec2a}

Figure \ref{Fig1}(a,d) shows the hexagonal real space for the
monolayer graphene with two inequivalent atoms in the unit cell. The
origin is set at the highest symmetry point, i.e. at the center of a
hexagon. The reciprocal space is shown in Fig. \ref{Fig1}(g)
highlighting the high symmetry points $\Gamma$, K, K$^{\prime}$, M
and lines T, T$^{\prime}$, $\Sigma$. Any other generic point outside
the high symmetry lines and points is named here \emph{u}. The
monolayer graphene on an isotropic medium has the space group
P6/$mmm$ (D$_{6h}^{1}$) in the Hermann-Mauguin notation. At the
$\Gamma$ point, the group of wavevector (GWV) is isomorphic to the
point group D$_{6h}$ (the Schoenflies character tables for the point
groups can be found in Ref. \cite{adobook}).

\begingroup
\begin{figure*}
\centering
\includegraphics [scale=0.5]{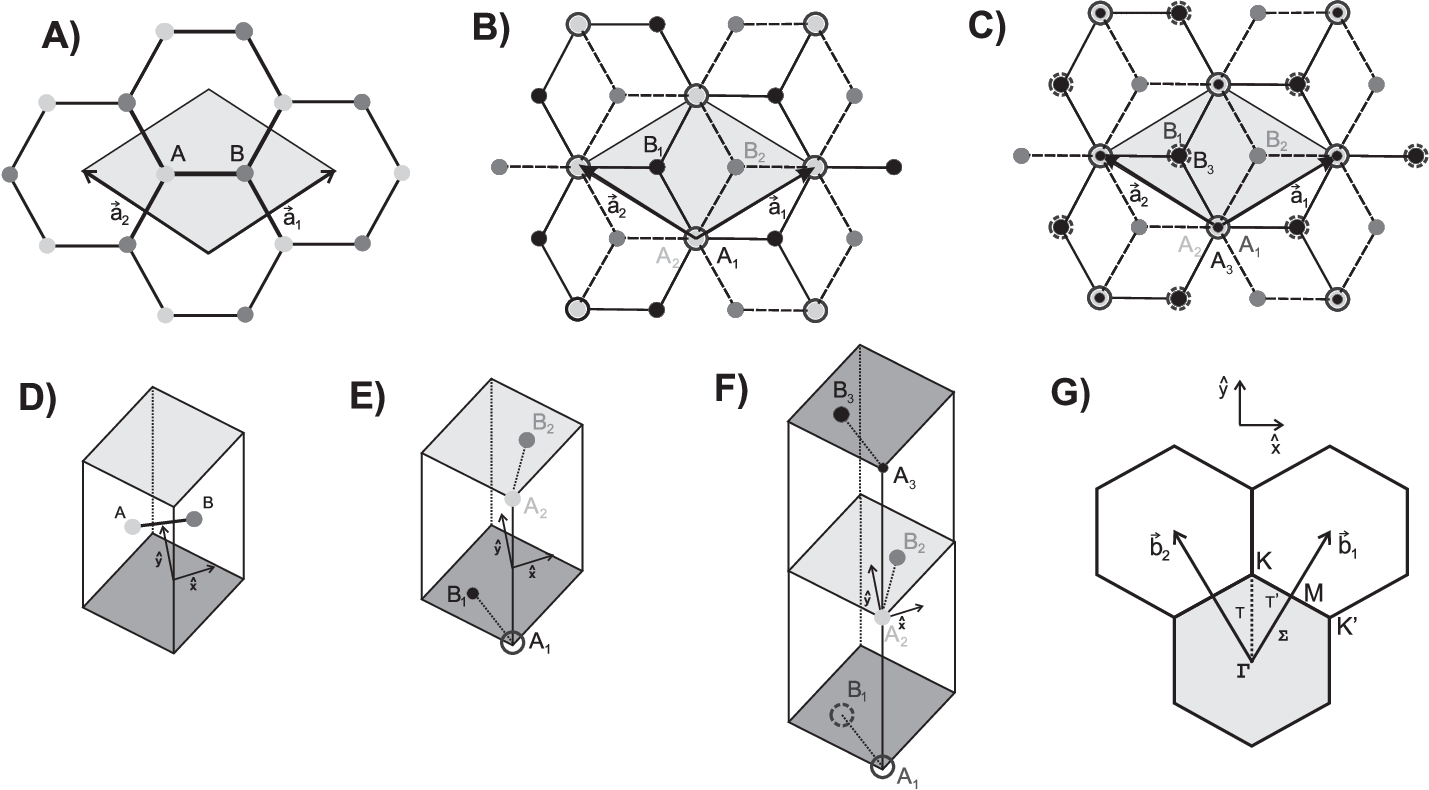}
\caption{\label{Fig1}(a) The real space top-view of a monolayer
graphene showing the non-equivalent A and B atoms. (b) The real
space top-view of bilayer graphene. Light and dark gray dots, and
black circles and dots represent the atoms in the upper and lower
layers, respectively. (c) The real top-view trilayer graphene. Small
and large black dots, light and dark gray dots, and traced and solid
circles represent the atoms in the upper, middle and lower layers,
respectively. The unit cell of the (d) monolayer, (e) bilayer and
(f) trilayer graphene. (g) The reciprocal space showing the 1$^{st}$
Brillouin zone in light gray, the high symmetry points and lines and
the two primitive vectors.}
\end{figure*}
\endgroup

The real space for bilayer and trilayer graphene with AB Bernal
stacking are show in Figs. \ref{Fig1}(b,e) and (c,f), respectively.
The symmetries for N-layer graphene, with N even or odd (from now
on, N$\neq1$), are the same of bilayer and trilayer graphene,
respectively. The main symmetry operation distinguishing the point
groups between even and odd layers are the horizontal mirror plane,
which is absent for N even, and the inversion, which is absent for N
odd. The point groups isomorphic to the GWV for mono-, N-layer
graphene (N even and odd), and for N infinite (graphite) are listed
in table\,\ref{tableprincipal3} for all points and lines in the
first Brillouin Zone (BZ). The GWV for N-layers graphene are
subgroups of the GWV for single layer graphene. The direct product
between the group from N even and N odd gives the graphene GWV, i.e.
$$\{G_{even}|0\} \otimes \{G_{odd}|0\}=\{G_{monolayer}|0\}.$$ On
graphite, the wavevector point groups are isomorphic to the
wavevector point groups of monolayer graphene, but differ
fundamentally for some classes where a translation of $c/2$ is
present, graphite belonging to the P6$_3$/$mmc$ (D$_{6h}^{4}$)
non-symorphic space group.

\begin{table}[htbp]
\begin{footnotesize}
\caption
  {The space groups and wavevector point groups for mono-, N-layer graphene and graphite at all points in the BZ.}
 \label{tableprincipal3}
  \centering
  \begin{tabular}{|c|c|c|c|c|c|c|c|}
  \hline
   & Space group& $\Gamma$ & K (K$^{\prime}$) & M & T (T$^{\prime}$) & $\Sigma$ &
   u \\
    \hline
    Monolayer & P6/$mmm$ &  D$_{6h}$ &  D$_{3h}$ &  D$_{2h}$ & C$_{2v}$ & C$_{2v}$ & C$_{1h}$\\
    \hline
    N even & P$\overline{3}$$m1$ &  D$_{3d}$ &  D$_{3}$ &  C$_{2h}$ & C$_{2}$ & C$_{1v}$ & C$_{1}$\\
    \hline
    N odd & P$\overline{6}$$m2$ & D$_{3h}$ &  C$_{3h}$ &  C$_{2v}$ & C$_{1h}$ & C$_{2v}$ & C$_{1h}$\\
    \hline
    N infinite & P6$_{3}$/$mmc$ & D$_{6h}$ &  D$_{3h}$ &  D$_{2h}$ & C$_{2v}$ & C$_{2v}$ & C$_{1h}$\\
    \hline
  \end{tabular}
  \end{footnotesize}
\end{table}

\subsection{Lattice vibrations and $\pi$ electrons}

The representations for the lattice vibration ($\Gamma_{lat.vib.}$)
and for the $\pi$ electrons ($\Gamma_{\pi}$) are given by
$\Gamma_{lat.vib.}=\Gamma^{eq}\otimes\Gamma^{vector}$ and
$\Gamma_{\pi}=\Gamma^{eq}\otimes\Gamma^{z}$, respectively, where
$\Gamma^{eq}$ is the atom equivalence representation,
$\Gamma^{vector}$ is the representation for the vectors $x$, $y$ and
$z$. For $\Gamma_{\pi}$ we used only $\Gamma^{z}$, which is the
irreducible representation for the vector $z$, since $\pi$ electrons
in graphene are formed by p$_{z}$ electronic orbitals. The results
for all points and lines in the first BZ for the $\Gamma_{lat.
vib.}$ are found in Table\,\ref{tableprincipal1} and for the
$\Gamma_{\pi}$ in Table\,\ref{tableprincipal2}.

Table \ref{tableprincipal2} shows that the $\pi$ electrons in
monolayer graphene are degenerated at the K (Dirac) point, as
obtained by theory \cite{saitobook}. Figures \ref{fig2}(a), (b) and
(c) show the electronic structure of a mono-, bi- and tri-layer
graphene, respectively calculated via density functional theory
(DFT) \cite{kohnsham,siesta96,siesta02}. The symmetry assignments of
the different electronic branches shown in Fig. \ref{fig2} were made
according to the DFT projected density of states.

\begingroup
\squeezetable
\begin{table*}[htbp]
\scriptsize \caption
  {The $\Gamma_{lat. vib.}$ wavevector point-group representations for mono- and N-layer graphene at all points in the BZ.}
 \label{tableprincipal1}
  \centering
  \begin{tabular}{|l|c|c|c|}
  \hline
    \textbf{} & \textbf{Monolayer} & \textbf{N even} & \textbf{N odd}\\
    \hline
    \textbf{$\Gamma$} &   $\Gamma^{-}_{2}+\Gamma^{-}_{5}+\Gamma^{+}_{4}+\Gamma^{+}_{6}$ & $N(\Gamma^{+}_{1}+\Gamma^{+}_{3}+\Gamma^{-}_{2}+\Gamma^{-}_{3})$ & $(N-1)\Gamma^{+}_{1}+(N+1)\Gamma^{-}_{2}+(N+1)\Gamma^{+}_{3}+(N-1)\Gamma^{-}_{3}$ \\
    \hline
    \textbf{K} & $K^{+}_{1}+K^{+}_{2}+K^{+}_{3}+K^{-}_{3}$ & $N(K_{1}+K_{2}+2K_{3})$ & $NK^{+}_{1}+NK^{-}_{1}+[f(N)+2]K^{+}_{2}+[f(N-2)]K^{+ \ast}_{2}$\\
               &    &   &  $+NK^{-}_{2}+(N-1)K^{- \ast}_{2}$ \footnote{Where $f(N)=\sum_{m=0}^{\infty}[\Theta(N-4m-2)+3\Theta(N-4m-4)]$, where $\Theta(x)$ is equal to 0 if $x<0$ and equal 1 otherwise.}\\
    \hline
    \textbf{M} & $M_{1}^{+}+M_{2}^{+}+M_{3}^{+}+M_{2}^{-}+M_{3}^{-}+M_{4}^{-}$ & $N(2M^{+}_{1}+M^{+}_{2}+M^{-}_{1}+2M^{-}_{2})$ & $2NM_{1}+(N-1)M_{2}+(N+1)M_{3}+2NM_{4}$ \\
    \hline
    \textbf{T(T$^{\prime}$)} & $2T_{1}+T_{2}+2T_{3}+T_{4}$ & $3N(T_{1}+T_{2})$ & $(3N+1)10T^{+}+(3N-1)T^{-}$ \\
    \hline
    \textbf{$\Sigma$} & $2\Sigma_{1}+2\Sigma_{3}+2\Sigma_{4}$ & $N(4\Sigma_{1}+2\Sigma_{2})$ & $2N\Sigma_{1}+(N-1)\Sigma_{2}+(N+1)\Sigma_{3}+2N\Sigma_{4}$ \\
    \hline
    \textbf{u} & $4u^{+}+2u^{-}$ & $6Nu$ & $(3N+1)u^{+}+(3N-1)u^{-}$ \\
    \hline
   \end{tabular}

\end{table*}
\endgroup

\begingroup
\squeezetable
\begin{table*}[htbp]
\begin{footnotesize}
\scriptsize \caption
  {The $\Gamma_{\pi}$ wavevector point-group representations for mono- and N-layer graphene at all points in the BZ.}
 \label{tableprincipal2}
  \centering
  \begin{tabular}{|l|c|c|c|}
  \hline
    \textbf{} & \textbf{Monolayer} & \textbf{N even} & \textbf{N odd}\\
    \hline
    \textbf{$\Gamma$} & $\Gamma^{-}_{2}+\Gamma^{+}_{4}$ & $N(\Gamma^{+}_{1}+\Gamma^{-}_{2})$ & $(N-1)\Gamma^{+}_{1}+(N+1)\Gamma^{-}_{2}$ \\
    \hline
    \textbf{K(K$^{\prime}$)} & $K^{-}_{3}$ & $\frac{N}{2}(K_{1}+K_{2}+K_{3})$ & $(\frac{N-1}{2})K^{+}_{1}+(\frac{N+1}{2})K^{-}_{1}+g(N)K^{+ \ast}_{2}(K^{+}_{2})+ g(N-2)(K^{+}_{2})(K^{+ \ast}_{2})+g(N)K^{-}_{2}+g(N+2)K^{-\ast}_{2}$ \footnote{Where $g(N)=\sum_{m=0}^{\infty}\Theta(N-4m-2)$,
where $\Theta(x)$ is equal to 0 if $x<0$ and equal 1 otherwise.}\\
    \hline
    \textbf{M} & $M_{3}^{+}+M_{2}^{-}$ & $N(M^{+}_{1}+M^{-}_{2})$ & $(N-1)M_{1}+(N+1)M_{4}$ \\
    \hline
    \textbf{T(T$^{\prime}$)} & $T_{2}+T_{4}$ & $N(T_{1}+T_{2})$ & $(N-1)T^{+}+(N+1)T^{-}$ \\
    \hline
    \textbf{$\Sigma$} & $2\Sigma_{4}$ & $2N\Sigma_{1}$ & $(N-1)\Sigma_{1}+(N+1)\Sigma_{4}$ \\
    \hline
    \textbf{u} & $2u^{-}$ & $2Nu$ & $(N-1)u^{+}+(N+1)u^{-}$ \\
    \hline
   \end{tabular}
  \end{footnotesize}
\end{table*}
\endgroup

\begingroup
\begin{figure*}
\centering
\includegraphics [scale=0.4]{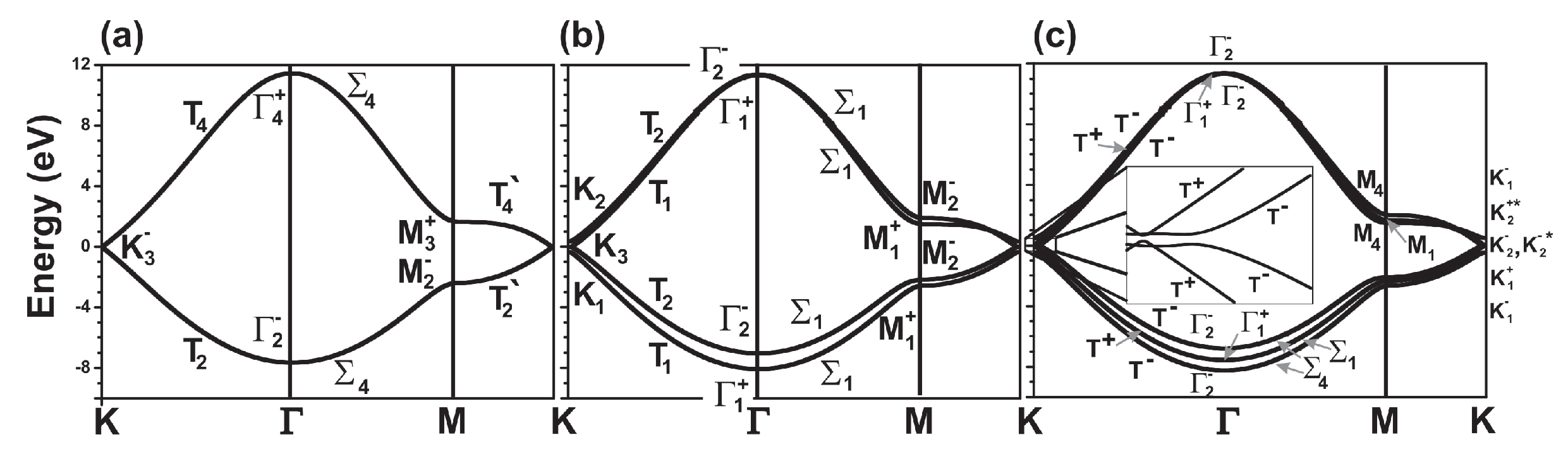}
\caption{\label{fig2} The electronic dispersion for the $\pi$
electrons calculated by DFT and the irreducible representations
($\Gamma_{\pi}$) for (a) monolayer, (b) bilayer and (c) trilayer
along the K$\Gamma$MK directions. The calculation was done via
pseudopotencial DFT \cite{kohnsham} as implemented in the SIESTA
program \cite{siesta96, siesta02}. We used a basis set composed of
pseudo atomic orbitals of finite range and the Local Density
Approximation (LDA) with the Ceperley-Alder parametrizationfor the
exchange-correlation functional.}
\end{figure*}
\endgroup

The bilayer graphene with AB Bernal stacking [see
Fig.\,\ref{Fig1}(b,e)] is also a zero gap semiconductor composed by
two conduction and two valence bands, and the electrons exhibit a
parabolic dispersion near the K point. Two bands are degenerated at
the K point (see Table\,\ref{tableprincipal2} and Fig.
\ref{fig2}(b)) and the other two have a gap of $2\gamma_{1}$, where
$\gamma_{1}$ is the Slonczewski-Weiss-McClure parameter
\cite{McClure1957, SW1958} that have a experimental value of $\sim$
0.3 - 0.4eV \cite{ohta06,ohta2007,lmalard07}.

Trilayer graphene in the ABA Bernal stacking (see Fig.
\ref{Fig1}(c,f)), belongs to the D$_{3h}$ point group and Fig.
\ref{fig2}(c) shows its electronic dispersion. The K point of
trilayer is isomorphic to $C_{3h}$. In Tables \ref{tableprincipal1}
and \ref{tableprincipal2}, $K^{+}_{2}$ and $K^{+*}_{2}$ are the two
one-dimensional representations of the $K^{+}_{2}$ representation,
where $\ast$ means the complex conjugate. The same happens for the
$K^{-}_{2}$ representation. The electron representations will be
given by $\Gamma_{\pi}^{K}=K^{+}_{1}+2K^{-}_{1}+K^{+
*}_{2}+K^{-}_{2}+K^{- \ast}_{2}$ for the K point and
$\Gamma_{\pi}^{K^{\prime}}=K^{+}_{1}+2K^{-}_{1}+K^{+}_{2}+K^{-
\ast}_{2}+K^{-}_{2}$ for K$^{\prime}$ point. Although time reversion
symmetry can imply degenerency between complex conjugate
representations in ciclic groups, in graphene the complex
conjugation also takes K into K$^{\prime}$ point and, consequently,
there are no degenerated bands at the K (K$^{\prime}$) point, in
agreement with \emph{tight-binding} calculations when including the
$\gamma_{2}$ and $\gamma_{5}$ next-nearest-layer coupling parameters
\cite{tightbindingtri,mccann2008}. This energy gap is also obtained
from \emph{ab initio} calculations (see the inset of Fig. \ref{fig2}
(c) and Ref. \cite{latil2006}).

\subsection{Gated mono- and bi-layer graphene}

If the monolayer graphene is in the presence of a perpendicular
electric field (gated graphene), the Fermi level changes. The
presence of charge inhomogeneity caused by substrate and/or absorbed
water can generate the same effect as verified in transport
\cite{novoselovscience2004,zhang1} and Raman measurements
\cite{casiraghi2007}, where the Dirac point is shifted from the
neutrality point. In this case, the $\pi$ electrons loose the
horizontal mirror plane and the inversion symmetry, and the system
is isomorphic to the point group C$_{6v}$. The irreducible
representations for the $\Gamma_{\pi}$ for the gated graphene can be
found in Table\,\ref{tableprincipal4}. There is no gap opening at K
point for a perfect perpendicular electric field effect.

The biased bilayer graphene have attracted a lot of attention
recently because it is the only material known to have a tunable
energy gap \cite{ohta06,mccann2006,neto07biased,johan07biased},
promising for applications on devices and lasers with tunable
energy. The mechanism behind this feature is based on applying an
electric field perpendicular to the graphene layers, so that the two
layers will be under an inequivalent potential. Then it is possible
to open a gap at the K point, breaking the double degenerated
K$_{3}$ irreducible representation into two one-dimensional
irreducible representations. Since the biased bilayer graphene
brakes the inversion center symmetry, the group of the wavevector at
$\Gamma$ for perfect perpendicular electric field is isomorphic to
C$_{3v}$. Table\,\ref{tableprincipal4} shows that the biased bilayer
contains the two one-dimensional representations K$_{2}$ and
K$_{2}^{\ast}$ at the K point, then a gap opening is expected on the
basis of symmetry arguments of inequivalent layers.

The representations for the $\Gamma_{lat.vib.}$ of the gated
monolayer (or biased bilayer) are the same as the monolayer (or
bilayer) in an isotropic medium, given in table
\ref{tableprincipal1}. The electric field does not affect the
symmetries of the phonons.

\begingroup
\squeezetable
\begin{table}[htbp]
\begin{footnotesize}
\caption
  {The group of wavevector and its $\Gamma_{\pi}$ representations for gated monolayer and biased bilayer graphene.}
 \label{tableprincipal4}
  \centering
  \begin{tabular}{|l|c|c|c|c|}
  \hline
    \textbf{} & \multicolumn{2}{c|}{\textbf{Gated monolayer}} & \multicolumn{2}{c|}{\textbf{Biased Bilayer}}\\
    \hline
    {} & GWV & $\Gamma_{\pi}$ & GWV & $\Gamma_{\pi}$\\
    \hline
    \textbf{$\Gamma$} & C$_{6v}$ & $\Gamma_{1}+\Gamma_{4}$ & C$_{3v}$ & $4\Gamma_1$ \\
    \hline
    \textbf{K(K$^{\prime}$)} & C$_{3v}$ & $K_{3}$ & C$_3$ & $2K_{1}+K_{2}+K^{\ast}_{2}$ \\
    \hline
    \textbf{M} & $C_{2v}$ & $M_1+M_3$ & C$_{1v}$ & $4M_{1}$ \\
    \hline
    \textbf{T(T$^{\prime}$)} & C$_{1v}$ & $T_{1}+T_{2}$ & C$_1$ & $4T$ \\
    \hline
    \textbf{$\Sigma$} & C$_{1v}$ & $2\Sigma_{1}$ & C$_{1v}$ & $4\Sigma_{1}$ \\
    \hline
    \textbf{u} & C$_1$ & $2u$ & C$_1$ & $4u$ \\
    \hline
   \end{tabular}
  \end{footnotesize}
\end{table}
\endgroup

\section{Selection rules for electron-radiation interaction}\label{sec3}

The symmetry properties described in the previous section will now
be applied to physical processes. In this section we discuss the
selection rules for electron-radiation interaction in the dipole
approximation, with emphasis on the high symmetry lines T and
T$^{\prime}$ in the electronic dispersion, where interesting
phenomena occur.

\begingroup
\squeezetable
\begin{table}[htbp]
\footnotesize \caption
  {Selection rules for electron-radiation interaction with \emph{\^{x}} and \emph{\^{y}} light polarization in mono-, bi- and tri-layer (see Fig. \ref{Fig1} (g) for \emph{\^{x}} and \emph{\^{y}} definition). For N even and N odd the selection rules are the same as for bi- and tri-layer graphene, respectively.}
 \label{tableprincipal5}
  \centering
  \begin{tabular}{|l|c|c|l|}
  \hline
    {} & \textbf{BZ point} & \textbf{polarization} & \textbf{W(k)} \\
    \hline
    \textbf{monolayer} & T & $x$ $\in$ $T_{3}$ & $T_{2}\otimes T_{3}\otimes T_{4}$ non null \\
    {} & {} & $y$ $\in$ $T_{1}$ & $T_{2}\otimes T_{1}\otimes T_{4}$ null \\
    \cline{2-4}
    {} & \emph{u} & $x$, $y$ $\in$ $u^{+}$ & $u^{-}\otimes u^{+}\otimes u^{-}$ non null \\
    \hline
    \textbf{gated}  & T & $x$ $\in$ $T_{2}$ & $T_{1}\otimes T_{2}\otimes T_{2}$ non null \\
    {\textbf{monolayer}} & {} & $y$ $\in$ $T_{1}$ & $T_{1}\otimes T_{1}\otimes T_{2}$ null \\
    \cline{2-4}
    {} & \emph{u} & $x$, $y$ $\in$ $u$ & $u\otimes u\otimes u$ non null \\
    \hline
    \textbf{bilayer} & T & $x$ $\in$ $T_{2}$ & $T_{1}\otimes T_{2}\otimes T_{1}$ null \\
    {(N-even)} & {} & {} & $T_{1}\otimes T_{2}\otimes T_{2}$ non null \\
    {} & {} & {} & $T_{2}\otimes T_{2}\otimes T_{2}$ null \\
    {} & {} & $y$ $\in$ $T_{1}$ & $T_{1}\otimes T_{1}\otimes T_{1}$ non null \\
    {} & {} & {} & $T_{1}\otimes T_{1}\otimes T_{2}$ null \\
    {} & {} & {} & $T_{2}\otimes T_{1}\otimes T_{2}$ non null \\
    \cline{2-4}
    {} & \emph{u} & $x$, $y$ $\in$ $u$ & $u\otimes u\otimes u$ non null \\
    \hline
    \textbf{biased} & T & $x$, $y$ $\in$ $T$ & $T\otimes T\otimes T$ non null \\
    \textbf{bilayer}& & &\\
    \cline{2-4}
    {} & \emph{u} & $x$, $y$ $\in$ $u$ & $u\otimes u\otimes u$ non null \\
    \hline
    \textbf{trilayer} & T & $x$, $y$ $\in$ $T^{+}$ & $T^{+}\otimes T^{+}\otimes T^{+}$ non null \\
    {(N-odd)} & {} & {} & $T^{+}\otimes T^{+}\otimes T^{-}$ null \\
    {} & {} & {} & $T^{-}\otimes T^{+}\otimes T^{-}$ non null \\
    \cline{2-4}
    {} & \emph{u} & $x$, $y$ $\in$ $u^{+}$ & $u^{+}\otimes u^{+}\otimes u^{+}$ non null \\
    {} & {} & {} & $u^{+}\otimes u^{+}\otimes u^{-}$ null \\
    {} & {} & {} & $u^{-}\otimes u^{+}\otimes u^{-}$ non null \\
    \hline
   \end{tabular}
\end{table}
\endgroup

In the dipole approximation, the absorption of light in a material
is related to the wave functions of the electron states in the
valence ($\psi^{v}(\textbf{k})$) and conduction
($\psi^{c}(\textbf{k})$) bands and the polarization of the incoming
light ($\bf{P}$) by
$W(\textbf{k})\propto|\textbf{P}\cdot\langle\psi^{c}(\textbf{k})|\nabla|\psi^{v}(\textbf{k})\rangle|^{2}$
\cite{cardona,GruneisPRB2003}. Knowing the symmetry of the initial
an final states, and the representation that generates the basis
function of the light polarization vector (\emph{x, y} or \emph{z}),
group theory can be used to compute wether $W(\textbf{k})$ is null
or not. The results are summarized in Table\,\ref{tableprincipal5}
considering graphene layers laying in the ($x$,$y$) plane and light
propagating along $z$. In the case of graphene, the light absorption
up to $3$ eV occurs only at T, T$^{\prime}$ and \emph{u} points.

It is important to highlight some results given in Table
\ref{tableprincipal5}. In the case of monolayer graphene on an
isotropic medium, numerical calculations show an anisotropy in the
optical absorption \cite{GruneisPRB2003,cancado041,cancado042}. This
anisotropy has indeed a symmetry basis, as clearly seen when
analyzing the selection rules at the T line. Absorption by visible
light has to couple T$_{2}$ and T$_{4}$ $\pi$ electron symmetries
(see Fig.\,\ref{fig2}(a)). For the T line direction along
\emph{\^{y}}, the only allowed absorption is for light polarized
along the \emph{\^{x}} direction. For incident light polarization
along the \emph{\^{y}} direction, no absorption will occur along
K$\Gamma$ direction, giving rise to the optical absorption
anisotropy on graphene \cite{GruneisPRB2003,cancado041,cancado042}.
Outside the high symmetry T line there is a non-zero probability of
absorption and the anisotropy is obtained by defining orthogonal
basis, as shown in Ref. \cite{GruneisPRB2003}.

When the monolayer graphene is on top of a substrate, with the
influence of the environment changing the Fermi level, there will be
no change in the selection rules for electron-radiation interaction.
Along the T line, the $\pi$ electrons are described by T$_{1}$ and
T$_{2}$ representations, where T$_{2}$ and T$_{1}$ contain \emph{x}
and \emph{y} basis function, respectively. Again there will be no
absorption for \emph{y} polarization.

\begin{figure}
\includegraphics [scale=0.35]{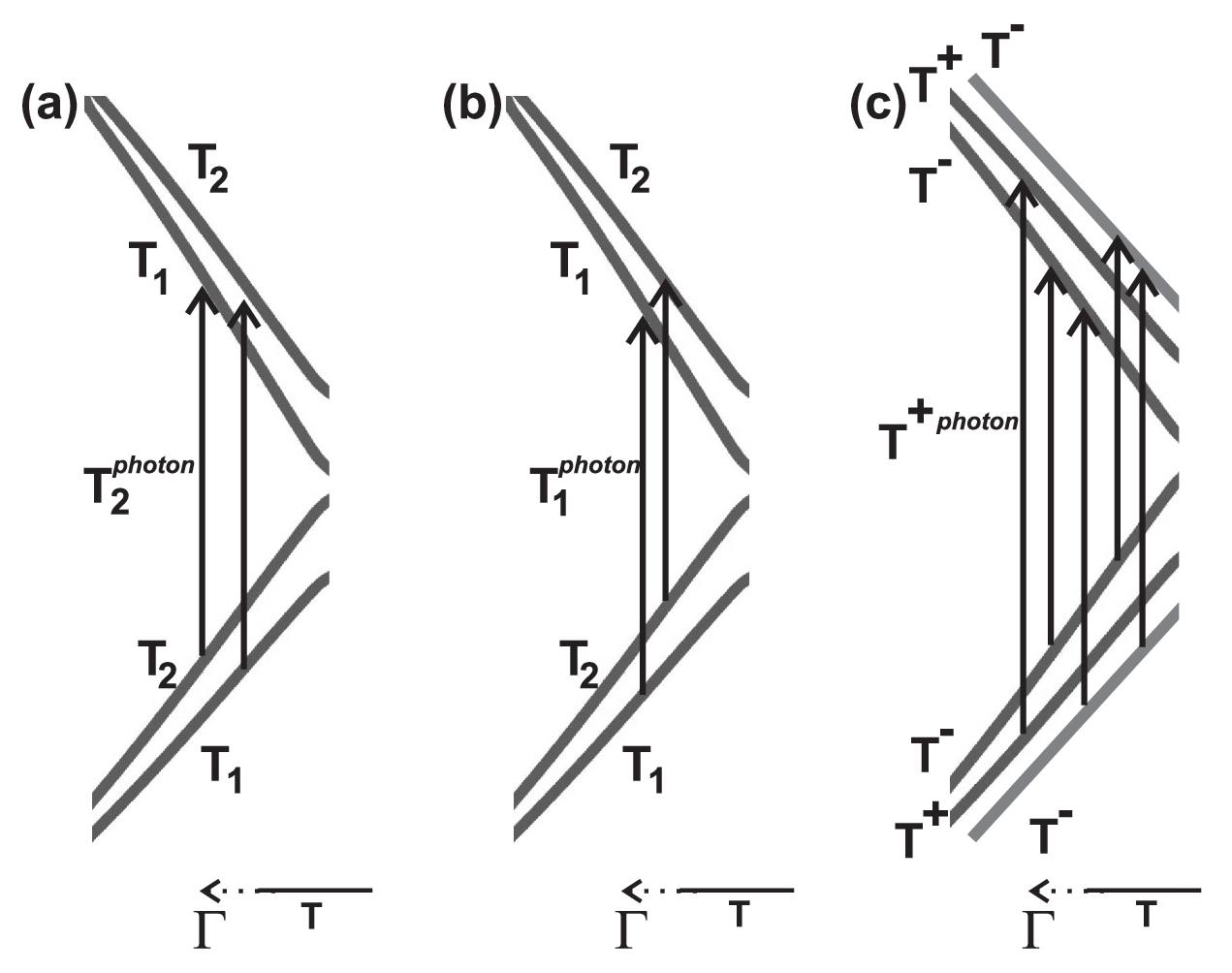}
\caption{\label{Fig3} (a,b) Schematic electron dispersion of
bi-layer graphene along the K$\Gamma$ direction showing the possible
transition induced by (a) a photon with T$_{2}$ symmetry (\emph{x}
polarization) and (b) a T$_{1}$ photon (\emph{y} polarization). (c)
The electronic dispersion of tri-layer graphene showing the five
possible transitions by light absorption.}
\end{figure}

The bilayer graphene is composed by four electronic bands at the T
line, belonging to two T$_{1}$ and two T$_{2}$ irreducible
representations. The four possible transitions are illustrated in
Fig.\,\ref{Fig3} (a,b). In this case both $x$ and $y$ polarized
light can be absorbed. For the biased bilayer graphene, all
electronic representations are the same, and it contains both $x,y$
base functions for light polarizations. Thus, all the four
transitions are allowed connecting all the four bands by the same
light polarization, differently from the unbiased bilayer case where
the light polarization selects the pair of bands that can be
connected.

The trilayer graphene will have more possibilities for light induced
transitions, since there are more possibilities between the three
$\pi$ and three $\pi^{*}$ bands. Along T(T$^{\prime}$) direction,
there are two T$^{+}$ and four T$^{-}$ bands giving raise to five
possible transitions (see Table \ref{tableprincipal5}), as shown in
Fig. \ref{Fig3} (c).

\section{Selection rules for the first-order Raman scattering and infrared absorption processes}\label{sec4}

The first-order Raman scattering process is limited to phonons at
the center of BZ ($\Gamma$ point) due to momentum conservation
requirement (phonon wavevector $q=0$). In monolayer graphene the
first-order Raman spectra is composed by the G band vibrational
mode, which is doubly degenerated at the $\Gamma$ point with
$\Gamma^{+}_{6}$ symmetry. The Raman active modes depending on N
(N$>$1) (without acoustic modes) are:

\begin{eqnarray*}
\Gamma^{\mathrm{Raman}}=N(\Gamma^{+}_{3}+\Gamma^{+}_{1}), \mathrm{\,\,for\,\,N \,\,even}\\
\Gamma^{\mathrm{Raman}}=N\Gamma^{+}_{3}+(N-1)(\Gamma^{-}_{3}+\Gamma^{+}_{1}),
\mathrm{\,\,for\,\,N\,\,odd}
\end{eqnarray*}

For even number of layers the G band belongs to the
$\Gamma^{+}_{3}$. There is a low frequency $\Gamma^{+}_{3}$ mode
with frequency depending on the number of layers (35-53
cm$^{-1}$)\cite{saha2008}. Two new Raman active modes near $\sim$80
cm$^{-1}$ and $\sim$ 900 cm$^{-1}$ appear with $\Gamma^{+}_{1}$
irreducible representations \cite{jiang2008,saha2008}. For odd
number of layers the G band is assigned as a combination of
$\Gamma_{3}^{+}$ and $\Gamma_{3}^{-}$ representations, also the
smaller wavenumber component is active in Raman by a
$\Gamma_{1}^{+}$ representation.

For monolayer graphene there is an infrared (IR) active mode
belonging to the $\Gamma_{2}^{-}$ representation, giving rise to an
absorption near $\sim$ 870 cm$^{-1}$. The IR active modes for N$>$1
are:

\begin{eqnarray*}
\Gamma^{\mathrm{IR}}=(N-1)\Gamma^{-}_{2}+(N-1)\Gamma^{-}_{3}, \mathrm{\,\,for\,\,N \,\,even}\\
\Gamma^{\mathrm{IR}}=N(\Gamma^{+}_{3}+(\Gamma^{-}_{2}),
\mathrm{\,\,for\,\,N\,\,odd}
\end{eqnarray*}

For even number of layers the active modes belong to the
$\Gamma_{2}^{-}$ and $\Gamma_{3}^{-}$ representations, the later one
referring to the $\sim$ 1590 cm$^{-1}$ frequency vibration
\cite{jiang2008,saha2008}. The infrared active modes for odd layer
number belong to $\Gamma_{2}^{-}$ and $\Gamma_{3}^{+}$ which are
also Raman active.

\section{Electron scattering by $q\neq0$ phonons}\label{sec5}

The electron-phonon scattering (EPS) is calculated from the initial
and final electron wave functions coupled by the phonon eigenvector
\cite{Jiang2005,netoeph2007} using the phonon-induced deformation
potencial. Therefore, the selection rules of the EPS processes are
obtained by the direct product of the symmetries of the initial and
final electronic states and the symmetry of the phonon involved in
the process. The allowed electron-phonon scattering processes for
monolayer, gated monolayer, bilayer, biased bilayer an trilayer
graphene along the K$\Gamma$ and KM directions (T and T$^{\prime}$,
lines respectively) and at a generic \emph{u} point are summarized
in Table \ref{tableprincipal6}.

\begingroup
\squeezetable
\begin{table}[h]
\begin{footnotesize}
\caption
  {Allowed processes for electron-phonon scattering for mono-, bi- and tri-layer graphene along the T and T$^{\prime}$ lines and at a generic \emph{u} point for each phonon symmetry. For N even and N odd the selection rules are the same as for bi- and tri-layer graphene, respectively.}
 \label{tableprincipal6}
  \centering
  \begin{tabular}{|l|c|c|l|}
  \hline
    {} & \textbf{BZ point} & \textbf{phonon} & \textbf{allowed scattering} \\
    \hline
    \textbf{monolayer} & T(T$^{\prime}$) & $T_{1}$ & $T_{2}\rightarrow T_{2}$, $T_{4}\rightarrow T_{4}$ \\
    {} & {} & $T_{3}$ & $T_{2}\rightarrow T_{4}$ \\
    \cline{2-4}
    {} & \emph{u} & $u^{+}$ & $u^{-}\rightarrow u^{-}$ \\
    \hline
    \textbf{gated} & T(T$^{\prime}$) & $T_{1}$ & $T_{1}\rightarrow T_{1}$, $T_{2}\rightarrow T_{2}$ \\
    \textbf{monolayer} & {} & $T_{2}$ & $T_{1}\rightarrow T_{2}$ \\
    \cline{2-4}
    {} & \emph{u} & $u$ & $u\rightarrow u$ \\
    \hline
   \textbf{bilayer} & T(T$^{\prime}$) & $T_{1}$ & $T_{1}\rightarrow T_{1}$, $T_{2}\rightarrow T_{2}$ \\
    {(N-even)} & {} & $T_{2}$ & $T_{1}\rightarrow T_{2}$ \\
    \cline{2-4}
    {} & \emph{u} & $u$ & $u\rightarrow u$ \\
    \hline
    \textbf{biased bilayer} & T(T$^{\prime}$) & $T$ & $T\rightarrow T$ \\
    \cline{2-4}
    {} & \emph{u} & $u$ & $u\rightarrow u$ \\
    \hline
    \textbf{trilayer} & T(T$^{\prime}$) & $T^{+}$ & $T^{+}\rightarrow T^{+}$, $T^{-}\rightarrow T^{-}$ \\
    {(N-odd)} & {} & $T^{-}$ & $T^{+}\rightarrow T^{-}$ \\
    \cline{2-4}
    {} & \emph{u} & $u^{+}$ & $u^{+}\rightarrow u^{+}$, $u^{-}\rightarrow u^{-}$ \\
    {} & {} & $u^{-}$ & $u^{+}\rightarrow u^{-}$ \\
    \hline
   \end{tabular}
  \end{footnotesize}
\end{table}
\endgroup

\section{Double resonance Raman scattering process}\label{sec6}

One example of explicit use of the electron-radiation and EPS
selection rules is the double resonance Raman scattering process
\cite{thomsen, saitoDR}, in which an electron in the conduction band
is scattered by a phonon with wavevector outside the $\Gamma$ point
in an intervalley (connecting electronic states near the K and
K$^{\prime}$ points) or in an intravalley (connecting electronic
state near the same K or K$^{\prime}$ point) process. The
G$^{\prime}$ Raman band ($\sim$ 2700 cm$^{-1}$) comes from an
intervalley process in which the electron is scattered by an
in-plane transversal optic (iTO) phonon. We will discuss in details
the G$^{\prime}$ scattering for mono- and multi-layer graphene.

\begin{figure}
\includegraphics [scale=0.30]{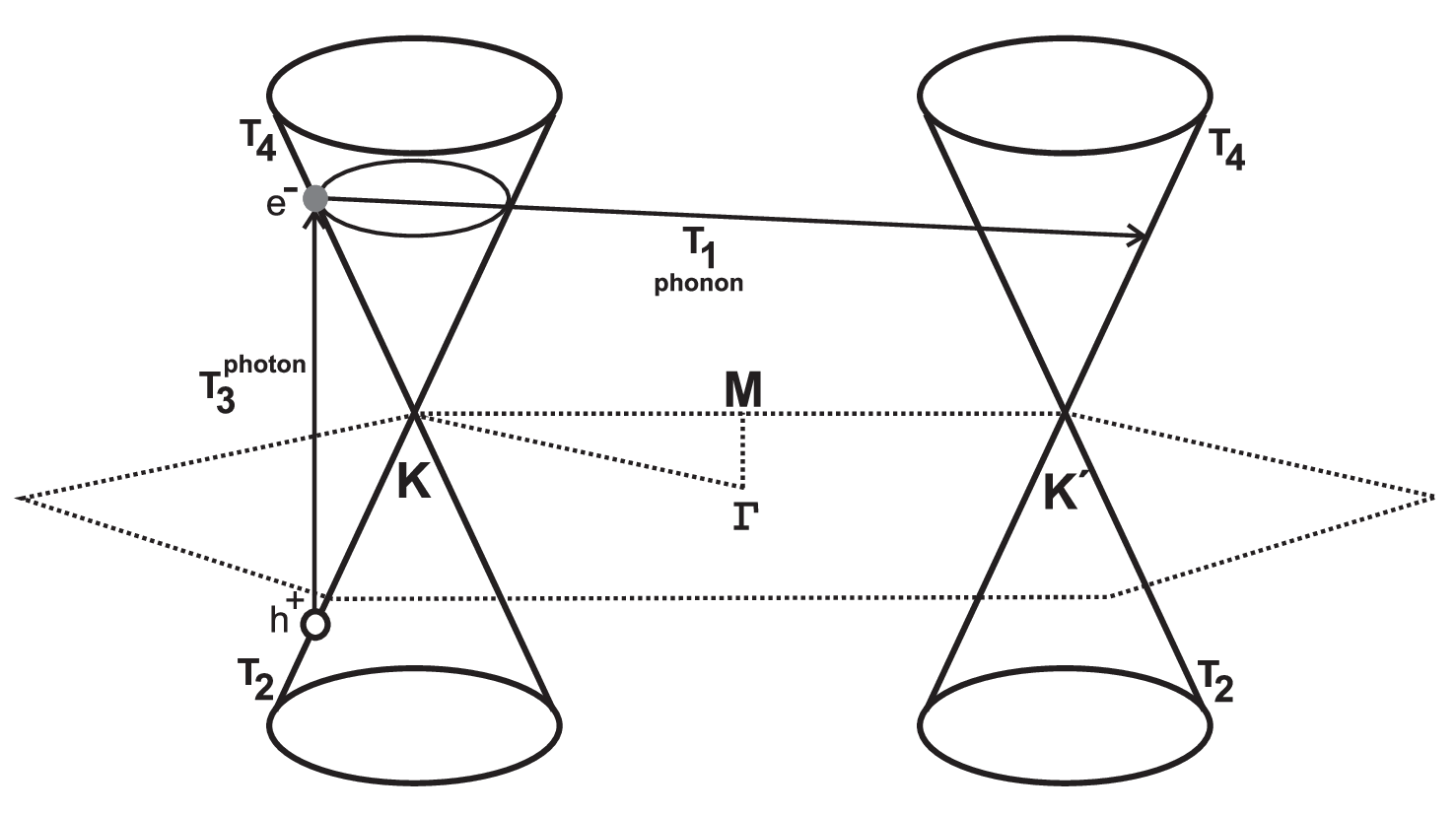}
\caption{\label{Fig4} The most efficient DRR process in graphene
showing the absorption of light with polarization symmetry T$_{3}$
followed by electron scattering by a phonon with T$_{1}$ symmetry.}
\end{figure}

For the monolayer graphene, the possible scattering is illustrate in
Fig. \ref{Fig4}. The iTO phonon at the KM (T$^{\prime}$) direction
presents a T$_{1}$ symmetry \cite{maultzsch04}, which can only
connect two electrons with the same symmetry. Many other similar
scattering events are allowed by symmetry, involving electron in the
K$\Gamma$ (T) direction or at any general $u$ point inside the
circle defined by the T$_{3}$ photon energy. However, the matrix
element has a strong angular dependence and the scattering is
dominated by the T electrons, as discussed in Ref. \cite{bob07}.
Therefore, the G$^{\prime}$ Raman band has only one peak, with full
width at half maximum (FWHM) of $\sim$ 24 cm$^{-1}$ (see Fig.
\ref{Fig6} (a)) \cite{ferrari,bob07}. For the graphene on top of a
substrate, the same selection rules apply, and the expected number
of G$^{\prime}$ peaks is the same as for the isolated monolayer
graphene on an isotropic medium.

\begin{figure}
\includegraphics [scale=0.30]{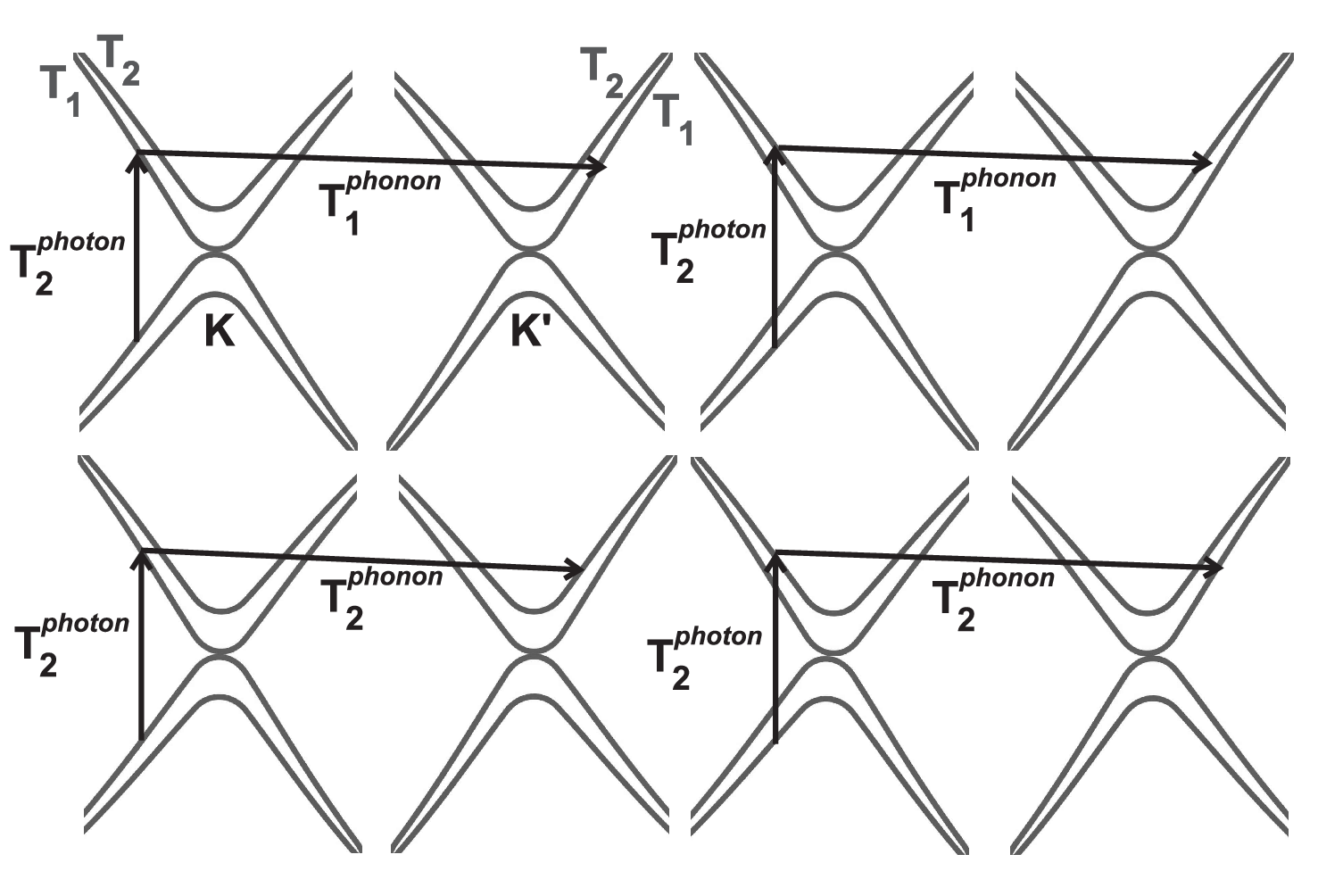}
\caption{\label{Fig5} The four allowed DRR processes in bilayer
graphene, taking into account the optical anisotropy.}
\end{figure}

\begin{figure}
\includegraphics [scale=0.6]{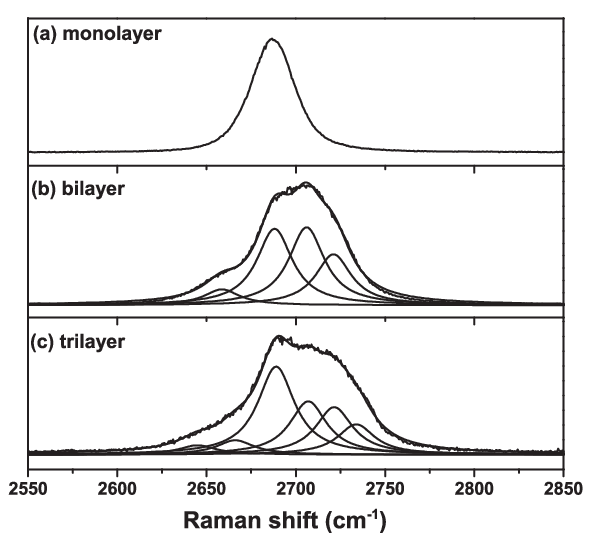}
\caption{\label{Fig6} The measured Raman spectra of the G$^{\prime}$
band of a (a) mono- (b) bi- and (c) trilayer graphene. The samples
were made by exfoliating graphite on top of a 100 nm silicon oxide
substrate using 2.41 eV laser energy. The G$^{\prime}$ band for
mono- bi- and tri-layer graphene were fitted with 1, 4 and 6
Lorentzians, respectively, with a FWHM of 24 cm$^{-1}$.}
\end{figure}

For the bilayer graphene, the number of allowed DRR processes
predicted by group theory will be larger, since both electronic and
phonon branches are doubled. Along the T line, there is polarization
dependence for the absorption linking different electronic bands, as
discussed in section \ref{sec3}. For wavevectors in the range of
visible light energy, the electron dispersion are almost linear,
then optical anisotropy can be applied here as for monolayer
graphene \cite{cancado2008}. Now, for computing the number of
resonant conditions involved in the DRR process, we are left with
only two excited electronic bands with symmetries T$_{1}$ and
T$_{2}$, which corresponds to Fig. \ref{Fig3} (a). The iTO phonons
for bilayer graphene have T$_1$ and T$_2$ symmetries. For the
electron scattering by a T$_{1}$ phonon, the allowed process are
between K and K$^{\prime}$ electronic bands with same symmetry
(T$_{1}$ $\rightarrow$ T$_{1}$ or T$_{2}$ $\rightarrow$ T$_{2}$).
The same happens with the electron scattering by a T$_{2}$ phonon,
but it connects conduction bands of different symmetries, i.e.
T$_{1}$ $\rightleftarrows$ T$_{2}$. This gives rise to four possible
DRR processes, as shown in Fig \ref{Fig5} \cite{ferrari}. The Raman
spectra can then be used to differentiate mono- and bi-layer
graphene (see Fig. \ref{Fig6})\cite{ferrari,gupta06,graf07}.

In the case of biased bilayer graphene, there are no selection rules
involving different photon polarizations. The biased bilayer can
have photon absorption linking all valence and conduction bands.
This leads to eight possible transitions considering EPS selection
rules for T$_{1}$ and T$_{2}$ phonons.

For the trilayer graphene, the DRR process will have again more
contributions because each phonon and electron band will be split in
three levels. Along the T line, there are five possibilities linking
the electronic bands between the K and K$^{\prime}$ points with a
T$^{+}$ phonon, and four possibilities for the T$^{-}$ phonon. The
total number of DRR process predicted by group theory will be
fifteen. However, the FWHM is large when compared to the energy
splitting between the G$^{\prime}$ Raman peaks, and when one makes
measurements of the G$^{\prime}$ Raman band, these fifteen peaks
cannot be distinguished, as illustrated in Fig. \ref{Fig6}. Similar
problem should happen for N$\geq$4.

\section{Summary}\label{sec7}

In this work we analyzed the symmetry aspects related to electrons
and phonons at each point in the BZ of graphene, depending on the
number of layers. The symmetry aspects can be generalized to any
value of N, differing for N even or odd. For monolayer and bilayer
we consider both an isotropic and an anisotropic medium. We derived
the selection rules for electron-radiation and electron-phonon
interactions. Some specific findings can be remarked:

$\bullet$ For the monolayer graphene, the predicted optical
anisotropy \cite{GruneisPRB2003} comes out directly from group
theory analysis. The electron-phonon scattering process is allowed
by symmetry at any generic point ($u$) in the Dirac cone, and the
observation of a single Lorentzian in the G$^{\prime}$ Raman band
comes from a strong anisotropy in the electron-phonon matrix element
\cite{bob07}.

$\bullet$ The gated graphene have lower symmetry, but the optical
anisotropy is still present, and for the DRR process, the symmetry
considerations are the same as graphene on an isotropic medium.

$\bullet$ In the case of bilayer graphene, the optical anisotropy is
also present and there are four dominant processes in the DRR. This
number increases to eight on biased bi-layer.

$\bullet$ In trilayer graphene, the number of possible DRR processes
is fifteen. However, the 15 processes are not distinguishable and
the G$^{\prime}$ Raman band can be nicely fit with 6 Lorentzians.
Similar situation is expected for larger number of layers.

\section*{Acknowledgements}

L.M.M, D.L.M and M.H.D.G contributed equally for this work and
acknowledge the Brazilian agency CNPq. This work was supported by
Rede Nacional de Pesquisa em Nanotubos de Carbono - MCT, FAPEMIG,
CNPq and Capes. We would like to thank M. A. Pimenta, L. G.
Can\c{c}ado, E. B. Barros and R. W. Nunes for useful discussions.

\appendix*
\section{Notation conversion from space group to point group irreducible
representations}\label{appendix}

In this work we derived the $\Gamma_{\pi}$ and $\Gamma_{lat.vib}$
for all points in the first BZ of multilayer graphene maintaining
the notation of space group (SG) for the irreducible
representations. The conversion to point group (PG) representation
is obtained considering that (\emph{\textbf{a}}) superscript sign
``+´´ or ``-´´ applies if the character of the horizontal mirror
($\sigma_{h}$) or inversion ($i$) is positive or negative,
respectively; (\emph{\textbf{b}}) the subscript number is given
following the order of the point group irreducible representations;
(\emph{\textbf{c}}) two representations can only have the same
number if they have superscript with positive or negative signs. As
an example we give in Table \ref{appendixtable} the $\Gamma$ point
space group notation conversion to the D$_{3h}$ (N-odd) and D$_{3d}$
(N-even) point groups and for the K point space group to the
C$_{3h}$ (N-odd) and D$_{3}$ (N-even) point groups.

\begingroup
\squeezetable
\begin{table}[htbp]
\begin{footnotesize}
\caption
  {Example of irreducible representation notation conversion from the $\Gamma$ point space group to D$_{3h}$ and D$_{3d}$ point groups, and from the K point space group to C$_{3h}$ and D$_{3}$ point groups.}
 \label{appendixtable}
  \centering
  \begin{tabular}{|c|c|c|c|c|c|c|c|}
  \hline
\multicolumn{4}{|c}{\textbf{$\Gamma$ point}}{}{} & \multicolumn{4}{|c|}{\textbf{K point}}{}{}\\
\hline
 \multicolumn{2}{|c|}{\textbf{D$_{3h}$}} & \multicolumn{2}{c|}{\textbf{D$_{3d}$}} & \multicolumn{2}{c|}{\textbf{C$_{3h}$}} & \multicolumn{2}{c|}{\textbf{D$_{3}$}}\\
 \hline
    \textbf{SG} & \textbf{PG} & \textbf{SG} & \textbf{PG} & \textbf{SG} & \textbf{PG} & \textbf{SG} & \textbf{PG}\\
    \hline
    $\Gamma_{1}^{+}$ & A$^{\prime}_{1}$ & $\Gamma_{1}^{+}$ & A$_{1g}$ & K$_{1}^{+}$& A$^{\prime}$ & K$_{1}$& A$_{1}$\\
    \hline
    $\Gamma_{1}^{-}$ & A$^{\prime\prime}_{1}$ & $\Gamma_{1}^{-}$ & A$_{1u}$ & K$_{1}^{-}$ & A$^{\prime\prime}$ & K$_{2}$& A$_{2}$\\
    \hline
    $\Gamma_{2}^{+}$ & A$^{\prime}_{2}$ & $\Gamma_{2}^{+}$ & A$_{2g}$ & K$_{2}^{+}$ & E$^{\prime}$ & K$_{3}$ & E \\
    \hline
    $\Gamma_{2}^{-}$ & A$^{\prime\prime}_{2}$ & $\Gamma_{2}^{-}$ & A$_{2u}$ & K$_{2}^{+\ast}$ & E$^{\prime\ast}$ & & \\
    \hline
    $\Gamma_{3}^{+}$ & E$^{\prime}$ & $\Gamma_{3}^{+}$ & E$_{g}$& K$_{2}^{-}$&E$^{\prime\prime}$ & & \\
    \hline
    $\Gamma_{3}^{-}$ & E$^{\prime\prime}$ & $\Gamma_{3}^{-}$ & E$_{u}$ & K$_{2}^{-\ast}$ & E$^{\prime\prime\ast}$ & & \\
    \hline
   \end{tabular}
  \end{footnotesize}
\end{table}
\endgroup

\end{document}